\documentclass[aps,prl,preprint,groupedaddress]{revtex4-1}

\usepackage{amsmath}

\begin{document}


\title{Eddy Current Model of Ball Lightning}


\author{J. D. Shelton}
\email[]{fixme@example.com}
\homepage[]{http://www.jasia.ca/jaydshelton/}
\affiliation{Independent Investigator}


\date{1993}

\begin{abstract}
Calculations show that high-energy ball lightning may consist of a ball of plasma containing a large circular electric current arising as an eddy current generated by lightning. Synthetic ball lightning might serve as a method of plasma confinement for purposes of nuclear fusion. In this paper, three articles concerning ball lightning and the related phenomenon of large ball lightning are combined to provide insight into this rarely glimpsed occurrence.
\end{abstract}


\maketitle

\section{}


\section{Introduction}
Calculations show that high-energy ball lightning may consist of a ball of plasma containing a large circular electric current arising as an eddy current generated by lightning. Synthetic ball lightning might serve as a method of plasma confinement for purposes of nuclear fusion. In this paper, three articles concerning ball lightning and the related phenomenon of large ball lightning are combined to provide insight into this rarely glimpsed occurrence.

Previous studies of ball lightning have concluded that no valid explanation of the phenomenon exists. This paper combines previous work with new insights to fashion a theory which accounts for the formation and some major properties of ball lightning.

To begin, calculations show that high-energy ball lightning may consist of a ball of plasma containing a large circular electric current arising as an eddy current generated in the lightning channel or from vortex formation within the channel.

According to the generally accepted theory of sunspots, the sun's differential rotation winds up magnetic field lines into cables, which are stable for years. There must be a large circular electric current inside a cable to generate the magnetic field. Ball lightning may just be a short length of magnetic cable and thus quite similar to a sunspot. Tornadoes have been reported to produce ball lightning, which is not surprising since tornadoes contain almost continuous internal lightning.

A simple ball of plasma at atmospheric pressure would be much lighter than air, would quickly cool and disperse and would display low energy. Ball lightning, however, evidently has a density comparable to air, is stable over a period of several seconds, and has appreciable energy, according to several reports. One must therefore postulate a mechanism which compresses and stabilizes the plasma.

One plausible mechanism is a large magnetic field, and observers' reports have corroborated magnetic effects associated with ball lightning. A magnetic field must arise from an electric current, and in a finite body an electric current must follow a closed path. One is thus led to suspect that ball lightning may be a ball of plasma containing a large circular electric current.

\section{The Model}

The plasma ball is formed when a lightning stroke induces a circular eddy current in a pocket of plasma adjacent to the lightning channel. As the eddy current absorbs energy from the stroke, it would necessarily exert a back-EMF on the channel, causing a voltage drop across the region. This ``bottleneck'' in the channel could cause a tremendous amount of energy to become concentrated in the region and transferred to the ball. Just as a large flow of water may result from the breakage of a dam in a river, an enormous current pulse could result when the lightning stroke forces its way past the eddy. A comparable current, approaching one million amperes or more, would be induced in the eddy. The radius of the ball would be comparable to the diameter of the lightning stroke, as one might expect if the currents are comparable. The plasma ball could persist after the lightning stroke has disappeared, if the large magnetic field compresses and stabilizes the plasma. The dense, hot plasma would also be highly opaque, preventing rapid energy loss.

It is often asserted (see Singer) that the virial theorem forbids the existence of a high-energy equilibrium ball of plasma. The assumption of equilibrium is required in the derivation of the virial theorem. Ball lightning (in this model), however, is not an equilibrium system. If one waits until equilibrium is reached, ball lightning ceases to exist. According to Landau and Lifshitz, the emission of radiation indicates that a system is not in equilibrium. Furthermore, an equilibrium system is characterized by periodic coordinates---the magnetic field and current in ball lightning certainly are not periodic. Ball lightning is an isolated system, but isolation is not equivalent to equilibrium. Even if a ball could be sustained by feeding energy into it, this system (like a waterfall) would be in a steady-state, but not in equilibrium, because the total entropy would continually increase. Hence, the virial theorem is inapplicable and nothing prevents the temporary existence of high-energy ball lightning---a conclusion which certainly agrees with observation. Similarly, the virial theorem does not forbid the existence of lightning, because lightning is also not an equilibrium system. The energy and duration of lightning are limited only by the available current. The same is true of ball lightning, which is essentially a circular form of lightning.

Shafranov used a hydrodynamic analogy to show that an external magnetic field is necessary to stabilize an equilibrium ball of plasma containing a ring current. The model presented herein, however, is not a simple ring current, so Shafranov's theorem does not apply.

Previous investigators have assumed that both poloidal and toroidal currents would be necessary for stability using an analogy with a tokamak. The tokamak, however, has a central hole and coils whereas ball lightning does not. Therefore, a more realistic model is needed.

\section{Analysis}
It is well known that electric currents flow more easily along magnetic field lines in a plasma. Taylor recently showed ``relaxed'' plasma configurations in which the current and magnetic fields are parallel. Theoretically, Taylor said, such a configuration is ``self-generating'' by a process of magnetic reconnection. Therefore, a plasma ball in this model might quickly adjust itself so that the electric current follows the path of least resistance; that is, where the current density $J$ is parallel to the magnetic field $B$:

\begin{equation}
	\nabla \times (B/\mu) = J = gB
\end{equation}

This equation in general has solutions, although the scalar geometric parameter g may have to be a function of position to satisfy boundary conditions. In a sphere, the field and current are helical, like a loop of twisted rope, equivalent to a superposition of poloidal and toroidal components. Helical solutions also exist for a long cylinder of plasma, such as a lightning stroke. A ball arising as an eddy current alongside or around the stroke could then obtain its initial helical structure from the stroke. Because the current and magnetic field are everywhere parallel, the magnetic force $J \times B$ is identically zero everywhere, no matter how great the current. Particles, however, could still be trapped by spiralling in the magnetic field, which would reduce somewhat the magnetic permeability, $\mu$.

A ball might form in a pocket of plasma tangent to the stroke with its axis perpendicular to the axis of the stroke. It seems more likely, however, that a ball would form around the stroke with axes coincident. If the stroke itself had a helical field structure, both poloidal and toroidal currents would be efficiently induced in the ball. In this position a ball would also be an effective choke, allowing it to absorb a great deal of energy from the stroke. The induced current opposes any change in the stroke current. When the stroke current begins to decrease, the current in the ball would flow in the same direction as the stroke current.

Spatial deformation of a ball conserves the flux \O. The magnetic energy $\O^2$/2$L$ is constant during a deformation, because there is no $J \times B$ force to convert the magnetic energy to any other form. Therefore the inductance $L$ is constant. If $L$ changed significantly, a great deal of energy would be released, and the ball would lose plasma and rapidly disintegrate. This agrees with observation; it is consistently reported that balls remain nearly constant in size until disappearance (often by explosion).

The inductive life span of a ball is given by

\begin{equation}
	\mu \hspace{4pt} \sigma \hspace{4pt} r^2
\end{equation}

where $\mu$ is the magnetic permeability, $\sigma$ is the conductivity, and $r$ is the radius of the ball. It is shown in kinetic theory that;

\begin{equation}
	\sigma = 2e^2/amv = 2e^2/(a\sqrt{3mkT})
\end{equation}

where $a$ is the cross-section of an ion; $m, v, e$ the mass, velocity, and charge of an electron; $k$ Boltzmann's constant; and $T$ temperature. According to the CRC tables, the radius of N$^+$ ion is 0.25\AA, and O$^+$ is 0.22\AA. Assuming an average of 0.24\AA$ $ and a temperature of at least 15,000K (the same as lightning) the conductivity is found to be approximately $10^8$ mhos/m (which is about the same as that of copper). This gives a lifetime of about one second, which agrees with observation. The quantum mechanical cross-section falls off as the inverse square of the temperature, so the resistance can be made arbitrarily low by choosing a sufficiently high temperature. A ball will be hotter inside, with 15,000K an estimate of the mean. If the temperature falls, the resistance would increase, additional ohmic heating would then drive up the temperature, and vice versa, so that a balance results.

Singer has estimated that the radiative life span of a ball of plasma at atmospheric pressure is about one millisecond. A ball of plasma at atmospheric density, however, would possess about 200 times as much energy, and the opacity would be much greater, resulting in a radiative life span measured in seconds, which agrees with observation. Hence, the life span of a ball of lightning is essentially the inductive or radiative lifetime, both of which are proportional to the square of the radius. The observed luminosity would be produced by an effective surface temperature of about 2,000K. The opacity given in published tables is consistent with the observed luminosity, which is inversely proportional to the opacity.

If diatomic air is heated to 15,000K (roughly 50 times room temperature) and broken into singly ionized ions and electrons, there are four times as many particles as before. Hence, a pressure of $4 \times 50 = 200$ atmospheres would result. If the energy of the system is a minimum with respect to the particle density, it can be shown that the magnetic permeability is two-thirds of the vacuum value, and the pressure is equal to the vacuum magnetic energy density:

\begin{equation}
B^2/2\mu_o = 200 = 200 \hspace{4pt} atmospheres = 2 \times 10^7N/m^2.
\end{equation}

This implies that $B$ is 6 tesla, and from the approximate equation:

\begin{equation}
B = \mu I/2r
\end{equation}

one can deduce that the current $I$ is about one million amperes. As a ball ages, the field weakens and plasma might escape, lightening the ball and forming rays or jets. If a mass m of n moles of plasma leaks and expands adiabatically, most of its internal energy would be converted to kinetic energy:

\begin{align}
\frac{1}{2} mv _e\hspace{2pt}^2 = (3/2) nRT \\
v_e = \sqrt{3RT / (m/n)} = \\
\sqrt{3\times 8 \times 15.000 / 0.007} = 7.000 m/sec
\end{align}

where we take the mass per mole of ionized nitrogen to be 0.007kg, $R$ the ideal gas constant, and $T$ the temperature. Conservation of momentum in the ball's instantaneous restframe requires:

\begin{align}
mdv = v_edm_o \\
v = v_eln(m_o/m)
\end{align}

where $v$ is the velocity of the ball, $m_o$ is the initial mass of the ball, and $m$ is the mass remaining after some is ejected at exhaust velocity $v_e$. Since the plasma is very hot and under great pressure, a jet of escaping plasma would act as a sort of rocket propulsion. Balls have also been seen to fall to the ground, roll about, and then rise, suggesting their density decreases.

The energy of ionization in a 0.1 meter ball at atmospheric density is about $4 \times 10^5$ joules; the magnetic energy, $\frac{1}{2}LI^2$, is about $10^5$ joules; and the power emitted would be about $10^5$ watts, which agrees with high-energy observations.

In a classic observation recorded by Singer, a ball exploded, demolishing a house. The energy of the ball was estimated to be greater than $10^9$ joules. A ball of radius 0.15 meter and current of 50 million amperes would have the density of water and energy of about $10^9$ joules. Singer also notes that balls have been reported to sink in water. Thus our model can account for the most extreme observations. Admittedly these figures seem rather high, but other models are even less plausible, or fail completely.

Balls often make a hissing or crackling sound, as one might expect in view of their electrical nature. The smell of sulphur so often reported is probably just ozone or oxides of nitrogen, copiously produced by the electrical discharge. The large current could easily kill, which has occurred. A ball need not carry any net charge, and the internal potential is only about one volt, which explains why balls show little interest in grounded conductors. Balls should be strongly magnetic, however, which has been verified. As the magnetic field decreases, a ball might become unstable and explode, which also is often observed. Explosion suggests a large internal pressure, consistent with our model.

\section{Large Ball Lighting \& UFOs}
Large ball lightning might explain some UFO reports. Other types of UFO might exist of course, even though many reports are undoubtedly hoaxes or misperceptions. It is well-known that large electric currents exist in the ionosphere. Suppose that a large specimen of ball lightning 10m in radius was created by such a current. It could weigh several tons, fall to the earth and burn a spot on the ground. It might live for several hours, but eventually its density would decrease and it would rise or hover. Jets of escaping plasma could disturb the earth beneath a ball, propel it to high velocity or impart rotation, which could result in an erratic or zigzag path.

The polar regions of a large ball would have less current density and hence might be lost, leaving a disk or lens-shaped object. A large ball, like lightning, would have a bright silvery-metallic appearance by day, and shine brightly by night. Irregularities in the field could produce light or dark spots on a ball; the color could vary, depending on the effective surface temperature. A row of evenly spaced lights might arise from a circular form of bead lightning, which probably consists of a standing longitudinal electric current wave. Intense heat, light and ultraviolet light would be radiated, as well as possibly radio interference or microwaves.

The current in a 10m ball at ground atmospheric density would be about $10^8$ amperes. The resulting large external magnetic field could disturb a compass up to a kilometer away, or open magnetically activated circuit breakers in nearby cars or electrical equipment. A large ball might appear to follow a car or airplane due to magnetic attraction to the iron, but large eddy currents induced in the metal would repel, preventing it from getting too close. Rotation of a large ball through its own magnetic field could produce a radial potential difference of thousands of volts, possibly giving rise to electrical effects such as beams, corona or electric shock. Perturbations by air currents or plasma jets could cause a large ball's magnetic axis to wobble about the direction of the earth's magnetic field, with a period of a few seconds. The net linear magnetic force on a ball is proportional to the gradient of the earth's field and is negligible.

A humming might be produced by radial oscillations of the ball. Pulsing or beeping could result from a periodic transfer of energy between different modes, or by atmospheric effects. Gases, such as ozone or nitrogen oxides, could be given off, producing a pungent smell causing nearby observers to become ill, incoherent or unconscious. This might explain odd stories related by witnesses.

\section{Conclusion}
Using this model, ball lightning, eddy currents and sunspots are basically the same, providing a much-needed unification of concepts. For example, luminous metal vapor spheres might be explained as specimens of ball lightning composed of metal vapor plasma rather than air plasma, because metal is more easily ionized than air. And, an electron might be represented in a classical model as a miniature globe of ball lightning.

This model explains the formation and demise, size, energy, life span, density, and other properties of ball lightning. Values of the properties that were derived agree with those observed in natural phenomena. The next theoretical step would be to calculate a detailed model of ball lightning, similar to a stellar structure. By creating a ball in the laboratory, one might determine if a ball could be sustained or raised to a higher temperature by feeding energy into it, for example by RF or particle beams. The ball could be held between two opposite magnetic poles, while gas jets or trimming fields are used to prevent it from touching the poles. Since natural ball lightning appears to be reasonably stable, a synthetic version might serve as a method of plasma confinement for purposes of nuclear fusion.

\section{References}
\begin{enumerate}
	\item Barry, James D. \emph{Ball Lightning and Bead Lightning.} New York: Plenum Press, 1980. 
	\item Blair, A.J.F. \emph{Nature.} 243, 1973: 512.
	\item Bruce, C.E.R. \emph{Nature.} 202, 1964: 996.
	\item Carpenter, D.G. \emph{Plasma Theory Applied to Ball Lightning.} Iowa State University Thesis No. 62-4145, 1962.
	\item ibid. \emph{AIAA Student Journal.} 1, 1963: 25.
	\item Eriksson, A.J. \emph{Nature.} 268, 1977: 35.
	\item Golde, R.H. \emph{Lightning.} New York, Academic Press, 1977.
	\item Jennison, R.C. \emph{Nature.} 245, 1973: 95.
	\item Johnson, P.O. American Journal of Physics. 33, 1965:119.
	\item Kitagawa, N. 1st International Conference on Ball Lightning (Tokyo). Singapore: World Scientific, 1989.
	\item Landau L.D. and E.M. Lifshitz. \emph{Electrodynamics of Continuous Media.} New York: Pergamon Press, 1960.
	\item ibid. \emph{The Classical Theory of Fields.} New York: Pergamon Press, 1962.
	\item Muhleisen, R. \emph{Kosmos.} 68, 1972: 159.
	\item Popov, Y.A. \emph{Priroda.} 48:12, 1959: 111.
	\item Shafranov, V.D. \emph{On Equilibrium Magnetohydrodynamic Configurations.} 3rd International Congress on Ionization Phenomena in Gases (Venice) 11-15 June 1957: 990-997.
	\item ibid. ``Soviet Physics.''; JETP. 6, 1957: 545.
	\item ibid. JETP. 10, 1960: 775.
	\item Shelton, J.D. \emph{Eddy Current Model of Ball Lightning.} Naturwissenschaften. 76, 1989: 373-374.
	\item ibid. \emph{Large Ball Lightning.} 
	\item ibid. U.S. Patent 4,654,561 (1987).
	\item Singer, Stanley. \emph{Ball Lightning: The Persistent Enigma.} Naturwissenschaften. 67, 1980: 332-337.
	\item ibid. \emph{The Nature of Ball Lightning.} New York: Plenum Press, 1971.
	\item Taylor, J.B. \emph{Relaxation and Magnetic Reconnection in Plasmas.} Reviews of Modern Physics. 58:3, July 1986.
	\item Viemeister, Peter E. \emph{The Lightning Book.} New York: Doubleday, 1961.
	\item Zou, Y.S. \emph{The Science of Ball Lightning.} 1st International Conference on Ball Lightning (Tokyo). Singapore: World Scientific. 1989: 273-287.
\end{enumerate}

\section{About the Author}
Jay Daniel Shelton attended the University of British Columbia, where he received a Masters degree in Physics.

Some of Jay's papers include: \emph{Eddy Current Model of Ball Lightning}, first published in Electric Spacecraft Journal, Issue 10, 1993 Leicester, NC 28748 USA; \emph{Quantum Gravity May Explain Dark Matter}, \emph{Rishon Model of Elementary Particles}, and \emph{Time Travel Model of Quantum Mechanics}.

Jay is an independent investigator and resides in Fruita, Colorado.

\end{document}